\documentclass[a4paper,twoside]{article}

\usepackage{listings}
\usepackage{epsfig}
\usepackage{subcaption}
\usepackage{calc}
\usepackage{amssymb}
\usepackage{amstext}
\usepackage{amsmath}
\usepackage{amsthm}
\usepackage{multicol}
\usepackage{pslatex}
\usepackage[bottom]{footmisc}
\usepackage{xspace}
\usepackage[usenames,dvipsnames]{xcolor}
\usepackage{adjustbox}
\usepackage{url}
\usepackage{cleveref}
\usepackage{tikz}
\usetikzlibrary{automata, arrows.meta, positioning,shapes.multipart}

\usepackage{wojtek15logics,wojtek15other}
\newtheorem{definition}{Definition}

\newcommand{\eg}{e.g.\xspace}
\newcommand{\ie}{i.e.\xspace}

\newcommand{\SAI}{SAI\xspace}

\newcommand{\AMAS}{AMAS\xspace}
\newcommand{\Agents}{\mathcal{A}} 
\newcommand{\A}{\ensuremath{A}} 
\newcommand{\Loc}{\ensuremath{L}}
\renewcommand{\loc}{\ensuremath{l}}
\newcommand{\Evt}{\ensuremath{Evt}}
\newcommand{\evt}{\ensuremath{e}}
\newcommand{\Prot}{\ensuremath{R}}
\newcommand{\Trans}{\ensuremath{T}}
\newcommand{\PV}{\mathit{PV}}
\renewcommand{\Val}{\ensuremath{V}}

\renewcommand{\model}{\mathit{M}}
\renewcommand{\States}{\ensuremath{S}}
\newcommand{\state}{\ensuremath{s}}

\newcommand{\seq}{\pi}
\newcommand{\strat}{\ensuremath{\sigma}}
\newcommand{\outcome}{\ensuremath{out}}

\usepackage{SCITEPRESS}

\begin{document}

\title{Towards Modelling and Verification of Social Explainable AI}
\author{\authorname{Damian Kurpiewski,\sup{1,3} Wojciech Jamroga,\sup{1,2} and Teofil Sidoruk\sup{1,4}}
\affiliation{\sup{1}Institute of Computer Science, Polish Academy of Sciences, Warsaw, Poland}
\affiliation{\sup{2}Interdisciplinary Centre for Security, Reliability, and Trust, SnT, University of Luxembourg}
\affiliation{\sup{3}Faculty of Mathematics and Computer Science, Nicolaus Copernicus University, Toruń, Poland}
\affiliation{\sup{4}Faculty of Mathematics and Information Science, Warsaw University of Technology, Warsaw, Poland}
\email{wojciech.jamroga@uni.lu, \{d.kurpiewski, t.sidoruk\}@ipipan.waw.pl}
}

\abstract{
Social Explainable AI (\SAI) is a new direction in artificial intelligence that emphasises decentralisation, transparency, social context, and focus on the human users.
\SAI research is still at an early stage. Consequently, it concentrates on delivering the intended functionalities, but largely ignores the possibility of unwelcome behaviours due to malicious or erroneous activity.
We propose that, in order to capture the breadth of relevant aspects, one can use models and logics of strategic ability, that have been developed in multi-agent systems.
Using the STV model checker, we take the first step towards the formal modelling and verification of \SAI environments, in particular of their resistance to various types of attacks by compromised AI modules.
}


\onecolumn \maketitle \normalsize \setcounter{footnote}{0} \vfill

\section{\uppercase{Introduction}}\label{sec:introduction}

Elements of artificial intelligence have become ubiquitous in daily life, being involved in social media, car navigation, recommender algorithms for music and films, and so on.
They also provide back-end solutions to many business processes, resulting in a huge societal and economical impact.
The idea of \emph{Social Explainable AI (\SAI)} represents an interesting new direction in artificial intelligence, which emphasises decentralisation, human-centricity, and explainability~\cite{SAI-website,Contucci22ai-ecosystem}.
This is in line with the trend to move away from classical, centralised machine learning, not only for purely technical reasons such as scalability constraints,
but also to meet the growing ethical expectations regarding transparency and trustworthiness of data storage and computation~\cite{Drainakis20federatedML,Ottun22socialFedLearn}.
The aim is also to put the human again in the spotlight, rather than concentrate on the technological infrastructure~\cite{Conti18internetOfPeople,Toprak21ego-arxiv,Fuchs22human-behavior-pt1-arxiv}.

\SAI is a new concept, and a subject of ongoing research. It still remains to be seen if it delivers AI solutions that are effective, transparent, and mindful of the user.
To this end, it should be extensively studied not only in the context of its intended properties, but also the possible side effects of interaction that involves AI components and human users in a complex environment.
In particular, we should carefully analyse the possibilities of adversarial misuse and abuse of the interaction, e.g., by means of impersonation or man-in-the-middle attacks~\cite{Dolev83security,Gollmann11security}.
In those scenarios, one or more nodes of the interaction network are taken over by a malicious party that tries to disrupt communication, corrupt data, and/or spread false information.
Clearly, the design of Social AI must be resistant to such abuse; otherwise it will be sooner or later exploited. 
While the topic of adversarial attacks on machine learning algorithms has recently become popular~\cite{Goodfellow18adversarialML,Kianpour19attacksML,Kumar20adversarialMLindustry}, 
the research on \SAI has been focused only on its expected functionalities.
This is probably because \SAI communities are bound to be conceptually, computationally, and socially complex.
A comprehensive study of their possible \emph{unintended} behaviors is a highly challenging task.

Here, we propose that \emph{formal methods for multi-agent systems}~\cite{Weiss99mas,Shoham09MAS} provide a good framework for multifaceted analysis of Social Explainable AI.
Moreover, we put forward a new methodology for such studies, based on the following hypotheses:
\begin{enumerate}
\item It is essential to formalise and evaluate multi-agent properties of \SAI environments. In particular, we must look at the properties of interaction between \SAI components that go beyond joint, fully orchestrated action towards a common predefined goal. This may include various relevant functionality and safety requirements. In particular, we should assess the impact of adversarial play on these requirements.
\item Many of those properties are underpinned by \emph{strategic ability} of agents and their groups to achieve their goals~\cite{Pauly02modal,Alur02ATL,Goranko15stratmas}. In particular, many functionality properties refer to the ability of legitimate users to complete their selected tasks. Conversely, safety and security requirements can be often phrased in terms of the inability of the ``bad guys'' to disrupt the behavior of the system.
\item Model checking~\cite{Clarke18principles} provides a well-defined formal framework for the analysis. Moreover, existing model checking tools for multi-agent systems, such as MCMAS~\cite{Lomuscio17mcmas} and STV~\cite{Kurpiewski21stv-demo} can be used to formally model, visualise, and analyse \SAI designs with respect to the relevant properties.
\item Conversely, \SAI can be used as a testbed for cutting-edge methods of model checking and their implementations.
\end{enumerate}

In the rest of this paper, we make the first step towards formal modelling, specification, and verification of \SAI.
We model \SAI by means of \emph{asynchronous multi-agent systems (\AMAS)}~\cite{Jamroga20POR-JAIR},
and formalise their properties using formulas of temporal-strategic logic \ATLs~\cite{Alur02ATL,Schobbens04ATL}.
For instance, one can specify that a malicious AI component can ensure that the remaining components will never be able to build a global model of desired quality, even if they all work together against the rogue element.
Alternatively, strategies of the ``good'' modules can be considered, in order to check whether a certain threshold of non-compromised agents is sufficient to prevent a specific type of attack.
Finally, we use the STV model checker~\cite{Kurpiewski21stv-demo} to verify the formalised properties against the constructed models. 
The verification is done by means of the technique of \emph{fixpoint approximation}, proposed and studied in~\cite{Jamroga19fixpApprox-aij}.

Note that this study does \emph{not} aim at focused in-depth verification of a specific machine learning procedure, like in~\cite{Wu20neural-verif-gamebased,Batten21neural-verif,Kouvaros21relu-neural-mcheck,Akintunde22neural-agents-mcheck}.
Our goal is to represent and analyse a broad spectrum of interactions, possibly at the price of abstraction that leaves many details out of the formal model.

The ideas, reported here, are still work in progress, and the results should be treated as preliminary.

\section{\uppercase{Social Explainable AI}}\label{sec:sai}

The framework of \emph{Social Explainable AI, or \SAI}~\cite{SAI-website,Contucci22ai-ecosystem,Fuchs22human-behavior-pt1-arxiv} aims to address several drawbacks inherent to the currently dominant AI paradigm.
In particular, today's state of the art machine learning (ML)-based AI systems are typically centralised.
The sheer scale of Big Data collections, as well as the complexity of deep neural networks that process them,
mean that effectively these AI systems act as opaque black boxes, non-interpretable even for experts.
This naturally raises issues of privacy and trustworthiness, further exacerbated by the fact that
storing an ever-increasing amount of sensitive data in a single, central location might eventually become unfeasible,
also for non-technical reasons such as local regulations regarding data ownership.

In contrast, \SAI envisions novel ML-based AI systems with a focus on the following aspects: 
\begin{itemize}
\item Individuation: a ``Personal AI Valet'' (PAIV) associated with each individual, acting as their proxy in a complex ecosystem of interacting PAIVs;
\item Personalisation: processing data by PAIVs via explainable AI models tailored to the specific characteristics of individuals;
\item Purposeful interaction: PAIVs build global AI models or make collective decisions starting from the local models by interacting with one another;
\item Human-centricity: AI algorithms and PAIV interactions driven by quantifiable models of the individual and social behaviour of their human users;
\item Explainability by design: extending ML techniques through quantifiable human behavioural models and network science analysis.
\end{itemize}

The current attempts at building \SAI use \emph{gossip learning} as the ML regime for PAIVs~\cite{Gossiping-website,Hegedus19gossip-learning,Hegedus21gossip-vs-federated-learning}.
An experimental simulation tool to assess the effectiveness of the process and functionality of the resulting AI components is available in~\cite{Lorenzo22sai-simulator}.
In this paper, we take a different path and focus on the multi-agent interaction in the learning process.
We model the network of PAIVs as an \emph{asynchronous multi-agent system (\AMAS)}, and formalise its properties as formulas of \emph{alternating-time temporal logic (\ATLs)}.
The formal framework is introduced in Section~\ref{sec:logic}. In Section~\ref{sec:models}, we present preliminary multi-agent models of \SAI, and show several attacks that can be modelled that way.
Finally, we formalise several properties, and conduct model checking experiments in Section~\ref{sec:experiments}.

\section{\uppercase{What Agents Can Achieve}}\label{sec:logic}

In this section, we introduce the formalism of Asynchronous Multi-agent Systems (\AMAS) \cite{Jamroga20POR-JAIR},
as well as the syntax and semantics of Alternating-time Temporal Logic \ATLs \cite{Alur02ATL,Schobbens04ATL},
which allows for specifying relevant properties of SAI models,
in particular the \emph{strategic ability} of agents to enforce a goal.

\subsection{Asynchronous MAS}

\AMAS can be thought of as networks of automata, where each component corresponds to a single agent.

\begin{definition}[\AMAS \cite{Jamroga20POR-JAIR}]\label{def:amas}
An \emph{asynchronous multi-agent system} (\AMAS) consists of $n$ agents $\Agents = \set{1,\dots,n}$,
each associated with a 7-tuple $\A_i = (\Loc_i, \iota_i, \Evt_i, \Prot_i, \Trans_i, \PV_i, \Val_i)$, where:
\begin{itemize}
\item $\Loc_i=\{\loc_i^1,\ldots,\loc_i^{n_i}\} \neq \emptyset$ is a finite set of \emph{local states};
\item $\iota_i \in \Loc_i$ is an \emph{initial local state};
\item $\Evt_i=\{\evt_i^1,\ldots,\evt_i^{m_i}\} \neq \emptyset$ a finite set of \emph{events};
\item $\Prot_i: \Loc_i \to 2^{\Evt_i}\setminus\{\emptyset\}$ is a \emph{repertoire of choices},
			assigning available subsets of events to local states;
\item $\Trans_i: \Loc_i \times \Evt_i \fpart \Loc_i$ is a (partial) \emph{local transition function}
			such that $\Trans_i(\loc_i,\evt)$ is defined iff $\evt\in \Prot_i(\loc_i)$.
			That is, $\Trans_i(\loc,\evt)$ indicates the result of executing event $\evt$ in state $\loc$ from the perspective of agent $i$;
\item $\PV_i$ is a set of the agent's \emph{local propositions},
			with $\PV_j$, $\PV_k$ (for $j \neq k \in \Agents)$ assumed to be disjoint;
\item $\Val_i: \Loc_i \then \powerset{\PV_i}$ is a \emph{valuation function}.
\end{itemize}
Furthermore, we denote:
\begin{itemize}
\item by $\Evt = \bigcup_{i \in \Agents} \Evt_i$, the set of all events;
\item by $\Loc = \bigcup_{i \in \Agents} \Loc_i$, the set of all local states;
\item by $Agent(\evt) = \set{i \in \Agents \mid \evt \in \Evt_i}$, the set of all agents which have event $\evt$ in their repertoires;
\item by $\PV = \bigcup_{i \in \Agents} \PV_i$ the set of all local propositions.
\end{itemize}
\end{definition}

The \emph{model} of an \AMAS provides its execution semantics with asynchronous interleaving of private events and synchronisation on shared ones.

\begin{definition}[Model]\label{def:model}
The \emph{model} of an \AMAS is a 5-tuple $\model = (\Agents, \States, \iota, \Trans, \Val)$, where:
\begin{itemize}
\item $\Agents$ is the set of \emph{agents};
\item $\States \subseteq \Loc_1\times\ldots\times\Loc_n$ is the set of \emph{global states},
			including all states reachable from $\iota$ by $\Trans$ (see below);
\item $\iota = (\iota_1,\dots,\iota_n) \in \States$ is the \emph{initial global state};
\item $\Trans: \States\times\Evt \fpart \States$ is the \emph{global transition function},
			defined by $\Trans(\state_1,\evt) = \state_2$ iff $\Trans_i(\state_1^i,\evt) = \state^i_2$ for all $i \in Agent(\evt)$
			and $\state_1^i = \state^i_2$ for all $i \in \Agents \setminus Agent(\evt)$
			(where $\state_j^i \in \Loc_i$ is agent $i$'s local component of $\state_j$);
\item $\Val: \States \rightarrow 2^{\PV}$ is the \emph{global valuation function},
			defined as $\Val(\loc_1,\dots,\loc_n) = \bigcup_{i\in\Agents} \Val_i(\loc_i)$.
\end{itemize}
\end{definition}

\subsection{Strategic Ability}

Linear and branching-time temporal logics, such as \LTL and \CTLs~\cite{Emerson90temporal}, have long been used in formal verification.
They enable to express properties about \emph{how} the state of the system will (or should) evolve over time.
However, in systems that involve autonomous agents,
whether representing human users or AI components
it is usually of interest \emph{who} can direct its evolution a particular way.

\ATLs \cite{Alur02ATL} extends temporal logics with \emph{strategic modalities} that allow for reasoning about such properties.
The operator $\coop{\A}\gamma$ says that agents in group (coalition) $\A$ have a \emph{strategy} to enforce property $\gamma$.
That is, as long as agents in $\A$ select events according to the strategy, $\gamma$ will hold no matter what the other agents do.
\ATLs has been one of the most important and popular agent logics in the last two decades.

\begin{definition}[Syntax of \ATLs]\label{def:syntax}
The language of \ATLs is defined by the grammar:
\begin{center}
$\varphi::= \prop{p} \mid \neg \varphi \mid \varphi\wedge\varphi \mid \coop{\A}\gamma$, \\
$\gamma::=\varphi \mid \neg\gamma \mid \gamma\land\gamma \mid \Next\gamma \mid \gamma\Until\gamma$,
\end{center}
where $\prop{p} \in \PV$ and $\A \subseteq \Agents$.
The definitions of Boolean connectives and temporal operators $\Next$ (``next'') and $\Until$ (``strong until'') are standard;
remaining operators $\Release$ (``release''), $\Always$ (``always''), and $\Sometm$ (``sometime'') can be derived as usual.
\end{definition}

Various types of strategies can be defined, based on the state information and memory of past states available to agents \cite{Schobbens04ATL}.
In this work, we focus on \emph{imperfect information, imperfect recall strategies}.

\begin{definition}[Strategy]
A \emph{memoryless imperfect information strategy} for agent $i \in \Agents$ is a function
$\strat_i \colon \Loc_i \to 2^{\Evt_i}\setminus\emptyset$ such that $\strat_i(\loc) \in \Prot_i(\loc)$ for each local state $\loc \in \Loc_i$.
A \emph{joint strategy} $\strat_\A$ of coalition $\A \subseteq \Agents$ is a tuple of strategies $\strat_i$, one for each agent $i \in \A$.
\end{definition}

The outcome set of a strategy collects all paths consistent with it,
\ie that may occur when the coalition follows the chosen strategy,
while opposing agents choose freely from their protocols.

\begin{definition}[Outcome]\label{def:outcome}
The \emph{outcome} of strategy $\strat_\A$ in global state $\state \in \States$ of model $\model$,
denoted by $\outcome_\model(\state,\strat_\A)$,
is the set of all paths $\seq = \state_0 \evt_0 \state_1 \evt_1 \dots$, such that
$\seq \in \outcome_\model(\state,\strat_\A)$
iff $\state_0 = \state$, and for each $j \geq 0$
$\begin{cases}
		\evt_j\in\strat_\A(\state_j^i) & \text{for every agent } i\in Agent(\evt_j)\cap \A, \\
		\evt_j\in\bigcup\Prot_i(\state_j^i) & \text{for every agent } i\in Agent(\evt_j)\setminus \A,
\end{cases}$

\noindent
where $\state^i$ denotes the local component $\loc_i$ of $\state$.
\end{definition}

\begin{definition}[Asynchronous semantics of \ATLs \cite{Jamroga20POR-JAIR}]
The asynchronous semantics of the strategic modality in \ATLs is defined by the following clause:
\vspace{0.1cm}

\noindent
$\model,\state \satisf \coop{\A}\gamma$ iff there is a strategy $\strat_\A$ such that $\outcome_\model(\state,\strat_\A) \neq \emptyset$ and,
for each path $\seq\in \outcome_\model(\state,\strat_\A)$, we have $\model,\seq \satisf \gamma$.
\vspace{0.1cm}

\noindent
The remaining clauses for temporal operators and Boolean connectives are standard, see~\cite{Emerson90temporal}.
\end{definition}

\section{\uppercase{Models}}\label{sec:models}

\begin{figure*}[!t]
\begin{tabular}{@{}cc@{}}
\hspace{-0.5cm}
\begin{tabular}{@{}c@{}}
  \begin{adjustbox}{width=1.2\columnwidth,left}
\begin{lstlisting}
Agent AI1:
init: start
%% ---Phase1: Gathering data---
start_gathering_data: start -> gather
gather_data: gather -[AI1_data<2]> gather 
        [AI1_data+=1]
%% 1: Incomplete data
stop_gathering_data: gather -[AI1_data < 1]> data_ready 
        [AI1_data=0, AI1_data_completion=1]
%% 2: Complete data
stop_gathering_data: gather -[1 <= AI1_data < 2]> data_ready 
        [AI1_data=0, AI1_data_completion=2]
%% 3: Too much data
stop_gathering_data: gather -[2 <= AI1_data]> data_ready 
        [AI1_data=0, AI1_data_completion=3]
skip_gathering_data: start -> data_ready
%% ---Phase2: Learning (building local model)---
start_learning: data_ready -> learn
keep_learning: learn -[AI1_information < 2]> learn 
        [AI1_information+=AI1_data_completion]
%% 1: Incomplete model
stop_learning: learn 
        -[AI1_information < 1 and AI1_model_quality > 0]> educated 
        [AI1_information=0, AI1_model_status=1, AI1_model_quality-=1]
stop_learning: learn 
        -[AI1_information < 1 and AI1_model_quality <= 0]> educated 
        [AI1_information=0, AI1_model_status=1]
%% 2: Complete model
stop_learning: learn 
        -[1 <= AI1_information < 2 and AI1_model_quality < 2]> educated 
        [AI1_information=0, AI1_model_status=2, AI1_model_quality+=1]
stop_learning: learn 
        -[1 <= AI1_information < 2 and AI1_model_quality >= 2]> educated 
        [AI1_information=0, AI1_model_status=2]
%% 3: Overtrained model
stop_learning: learn 
        -[2 <= AI1_information and AI1_model_quality > 0]> educated 
        [AI1_information=0, AI1_model_status=3, AI1_model_quality-=1]
stop_learning: learn 
        -[2 <= AI1_information and AI1_model_quality <= 0]> educated 
        [AI1_information=0, AI1_model_status=3]
skip_learning: data_ready -> sharing
%% ---Phase3: Sharing local models---
start_sharing: educated -> sharing
%% Share local model and get average quality of both models
%% receive left
shared share_3_with_1: sharing -> sharing2 
        [AI1_model_quality=%AI3_model_quality]
%% send right
shared share_1_with_2: sharing2 -> sharing3
end_sharing: sharing3 -> end
%% ---Phase4: End---
wait: end -> end
repeat: end -> learn
\end{lstlisting}
  \end{adjustbox}
\end{tabular}
 & 
\hspace{-7.7cm}
\begin{tabular}{@{}c@{}}
  \scalebox{0.55}{
    \newcommand{\hquad}{\hspace{0.5em}} 
\begin{tikzpicture}[>=stealth, node distance=2cm]
	\node[state] (start) {$q_0$};
	\node[state, right = 3cm of start] (gather) {$q_1$};
	\node[state, right = 6cm of gather] (data_ready) {$q_2$};
	\node[state, below = 7.5cm of data_ready] (learn) {$q_3$};
	\node[state, below = 7.5cm of start] (educated) {$q_4$};
	\node[state, below = 4.5cm of educated] (sharing) {$q_5$};
	\node[state, right = 3.5cm of sharing] (sharing2) {$q_6$};
	\node[state, right = 2.5cm of sharing2] (sharing3) {$q_7$};
	\node[state, below = 4.5cm of learn] (end) {$q_8$};

	%
	\path[->] (start) edge [above] node[align=center] {$start\_gathering$} (gather);
	\path[->] (gather) edge [bend left=60, above] node[align=center, sloped] {$\prop{[AI1\_data < 1]}$~$stop\_gathering$\qquad\qquad\qquad\qquad\\$\prop{AI1\_data=0, AI1\_completion=1}$\qquad\qquad} (data_ready);
	\path[->] (gather) edge [above] node[align=center, sloped] {$\prop{[1 <= AI1\_data < 2]}$~$stop\_gathering$\\$\prop{AI1\_data=0, AI1\_completion=2}$} (data_ready);
	\path[->] (gather) edge [bend right=60, below] node[align=center, sloped] {$\prop{[2 <= AI1\_data]}$~$stop\_gathering$\\$\prop{AI1\_data=0, AI1\_completion=3}$} (data_ready);
	\path[->] (start) edge [bend left=90, above] node[sloped] {$skip\_gathering$} (data_ready);
	\path[->] (data_ready) edge [above] node[align=center, sloped] {$start\_learning$} (learn);
	\path[->] (learn) edge [bend right=90, above] node[align=center, sloped]
		{$\prop{[AI1\_info < 1 \land AI1\_mqual > 0]}$~$stop\_learning$\\$\prop{AI1\_info=0, AI1\_mstatus=1, AI1\_mqual-=1}$} (educated);
	\path[->] (learn) edge [bend right=35, above] node[align=center, sloped]
		{$\prop{[AI1\_info < 1 \land AI1\_mqual \leq 0]}$~$stop\_learning$\\$\prop{AI1\_info=0, AI1\_mstatus=1}$} (educated);
	\path[->] (learn) edge [bend right=3, above] node[align=center, sloped]
		{$\prop{[1 \leq AI1\_info < 2 \land AI1\_mqual < 2]}$~$stop\_learning$\\$\prop{AI1\_info=0, AI1\_mstatus=2, AI1\_mqual+=1}$} (educated);
	\path[->] (learn) edge [bend left=3, below] node[align=center, sloped]
		{$\prop{[1 \leq AI1\_info < 2 \land AI1\_mqual \geq 2]}$~$stop\_learning$\\$\prop{AI1\_info=0, AI1\_mstatus=2}$} (educated);
	\path[->] (learn) edge [bend left=35, below] node[align=center, sloped]
		{$\prop{[2 \leq AI1\_info \land AI1\_mqual > 0]}$~$stop\_learning$\\$\prop{AI1\_info=0, AI1\_mstatus=3, AI1\_mqual-=1}$} (educated);
	\path[->] (learn) edge [bend left=90, below] node[align=center, sloped]
		{$\prop{[2 \leq AI1\_info \land AI1\_mqual \leq 0]}$~$stop\_learning$\\$\prop{AI1\_info=0, AI1\_mstatus=3}$} (educated);
	\path[->] (data_ready) edge [bend right=60, above] node[sloped] {$skip\_learning$} (sharing);
	\path[->] (educated) edge [above] node[align=center, sloped] {\qquad$start\_sharing$} (sharing);
	\path[->] (sharing) edge [below] node[align=center] {$share\_3\_with\_1$\\$\prop{AI1\_mqual=\%AI3\_mqual}$} (sharing2);
	\path[->] (sharing2) edge [below] node[align=center] {$share\_1\_with\_2$} (sharing3);
	\path[->] (sharing3) edge [below] node[align=center, sloped] {$end\_sharing$} (end);
	\path[->] (end) edge [above] node[align=center, sloped] {$repeat$} (learn);

	\path[->] (gather) edge[loop above] node[align=center] {$\prop{[AI1\_data<2]}$~$gather\_data$\qquad\qquad\qquad\qquad\\$\prop{AI1\_data+=1}$\qquad\qquad} (gather);
	\path[->] (learn) edge[loop right] node[align=center, sloped] {\quad$\prop{[AI1\_info<2]}$~$keep\_learning$\\$\prop{AI1\_info+=AI1\_completion}$} (learn);
	\path[->] (end) edge[loop right] node[align=center, sloped] {\hquad$wait$} (end);
\end{tikzpicture}
  }
\end{tabular}
\end{tabular}
\caption{Specification and graphical representation of the honest AI agent}
\label{fig:ai}
\label{fig:honest}
\end{figure*}

\begin{figure}[t]
\centering
\includegraphics[width=\columnwidth]{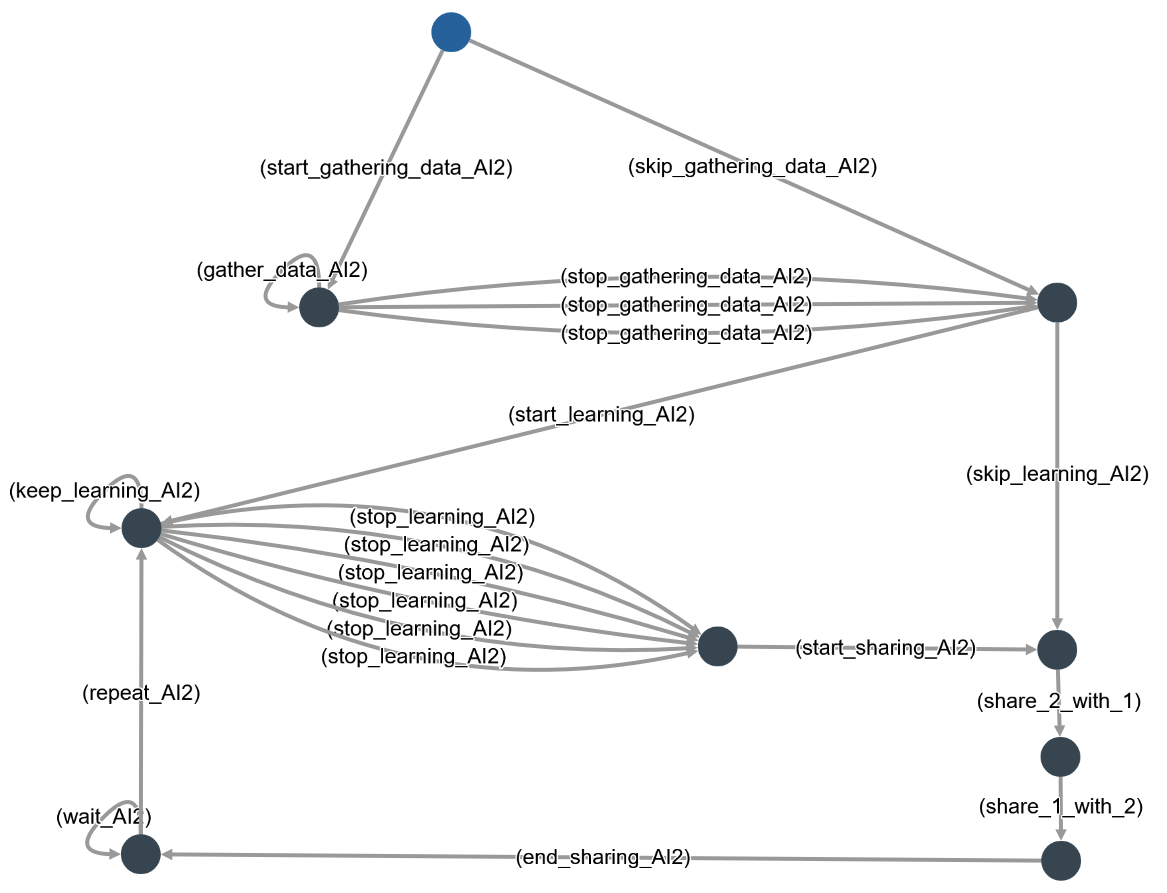}
\caption{Visualization of honest AI in STV\medskip}
\label{fig:honest-stv}
\end{figure}

The first step towards the verification of the interaction between agents in Social Explainable AI is a thorough and detailed analysis of the underlying protocol.
We begin by looking into the actions performed and the messages exchanged by the machines that take part in the learning phase.
Then, when we can grasp the whole overview of the system, we can start designing multi-agent models.
Usually, the system is too detailed to be modelled as it is.
In that case, we create the abstract view of the system.

\subsection{Agents}\label{sec:ai-agents}

In this work, we focus on the learning phase of the protocol.
We model each machine equipped with an AI module as a separate agent.
The local model of an AI agent consists of three phases: the data gathering phase, the learning phase and the sharing phase.

\para{Data gathering phase.}
In this phase, the agent is able to gather the data required for the learning phase.
The corresponding action can be performed multiple times, each time increasing the local variable that represents the amount of data gathered.
At the end of this phase, the amount of gathered data is analysed and, depending on the exact value, the agent's preparation is marked as incomplete, complete or too much data.
From this phase the agent goes directly to the learning phase.

\para{Learning phase.}
The agent can use the previously gathered data to train his local AI model.
The effectiveness of the training depends on the amount of data gathered.
Excessive data means that the model can be easily overtrained, while insufficient data may lead to more iterations required to properly train the model.
The training action can be performed multiple times each time increasing the local variable related to the quality of the model.
At the end of this phase the internal AI model can end to be overtrained, undertrained or properly trained.
After this phase the agent is required to share his model with other agents.

\begin{figure}
    \begin{adjustbox}{width=0.8\columnwidth,center}
\begin{lstlisting}
Agent Mim:
init: start
shared share_1_with_mim: start -> start 
    [Mim_model_quality=AI1_model_quality]

shared share_mim_with_1: start -> start

shared share_2_with_mim: start -> start 
    [Mim_model_quality=AI2_model_quality]

shared share_mim_with_2: start -> start
\end{lstlisting}
    \end{adjustbox}
    \caption{Specification of the Man in the Middle agent}
    \label{fig:mim}
\end{figure}

\begin{figure}[t]
  \centering\medskip
    \scalebox{0.75}{
      \begin{tikzpicture}[>=stealth, node distance=2cm]
	\node[state] (start) {$q_0$};

	\path[->] (start) edge[loop above] node[align=center] {$\prop{Mim\_model\_quality=AI1\_model\_quality}$\\$share\_1\_with\_mim$} (start);
	\path[->] (start) edge[loop below] node[align=center] {$share\_2\_with\_mim$\\$\prop{Mim\_model\_quality=AI2\_model\_quality}$} (start);
	\path[->] (start) edge[loop left] node {$share\_mim\_with\_1$} (start);
	\path[->] (start) edge[loop right] node {$share\_mim\_with\_2$} (start);
	
\end{tikzpicture}
    }
\caption{Graphical representation of Man in the Middle}
\label{fig:mim-amas}
\end{figure}

\para{Sharing phase.}
Agents share their local AI models with each other following a simple sharing protocol.
This protocol is based on the packet traversal in the ring topology.
Each agent receives the model from the agent with previous ID and sends his model to the agent with next ID, while the last agent shares his model with the first agent to close the ring.
In order to avoid any deadlocks, each agent with odd ID first receives the model and then sends his own, and each agent with even ID first sends his own model and then receives the model from the agent before him.

When receiving the model, the agent can either accept it or reject it, and his decision is based on the quality of the model being shared.
After accepting the model, the agent merges it with his own and the resulting model quality is the maximum of the two.

After the sharing phase, the agent can go back to the learning phase to further train his model.

To formalize the details of the procedure, we have utilised the open-source experimental model checker STV~\cite{Kurpiewski21stv-demo}, which was used \eg to model and verify the real-world voting protocol Selene~\cite{Selene}.
Figure~\ref{fig:honest} presents the STV code specifying the behavior of the honest AI component (left) and its detailed representation as an \AMAS (right).
Figure~\ref{fig:honest-stv} shows the visualization of the component, produced by the tool.

\subsection{Attacks}

The model described in Section~\ref{sec:ai-agents} reflects the ideal scenario in which each agent is honest and directly follows the implemented protocol.
Of course, it is known to not always be the case.
One of the machines can malfunction, which can result in taking actions not permitted in the protocol.
On the other hand, the malicious software can also infect one of the agents, forcing him to function improperly.
In our work we research two possible scenarios of the attack: the man in the middle attack and the impersonator attack.

\para{Man in the middle.}
Let us assume that there exists another, dishonest agent.
We will call him the intruder.
This agent does not participate in the data gathering or learning phases, but he is particularly interested in the sharing phase.
The intruder can intercept any model that is being sent by one of the honest agents and then pass it to any other agent.
The STV code for the man-in-the-middle attacker is presented in Figure~\ref{fig:mim}, and its graphical representation in Figure~\ref{fig:mim-amas}.

\begin{figure}
    \begin{adjustbox}{width=0.8\columnwidth,center}
\begin{lstlisting}
Agent AI2:
init: start
set_quality_0: start -> set_quality 
                [AI2_model_quality=0]

set_quality_1: start -> set_quality 
                [AI2_model_quality=1]

set_quality_2: start -> set_quality
                [AI2_model_quality=2]

shared share_2_with_3: set_quality -> sharing
shared share_1_with_2: sharing -> start
\end{lstlisting}
    \end{adjustbox}
    \caption{Specification of the Impersonator agent}
    \label{fig:imp}
\end{figure}

\begin{figure}[t]
\centering
\includegraphics[width=0.6\columnwidth]{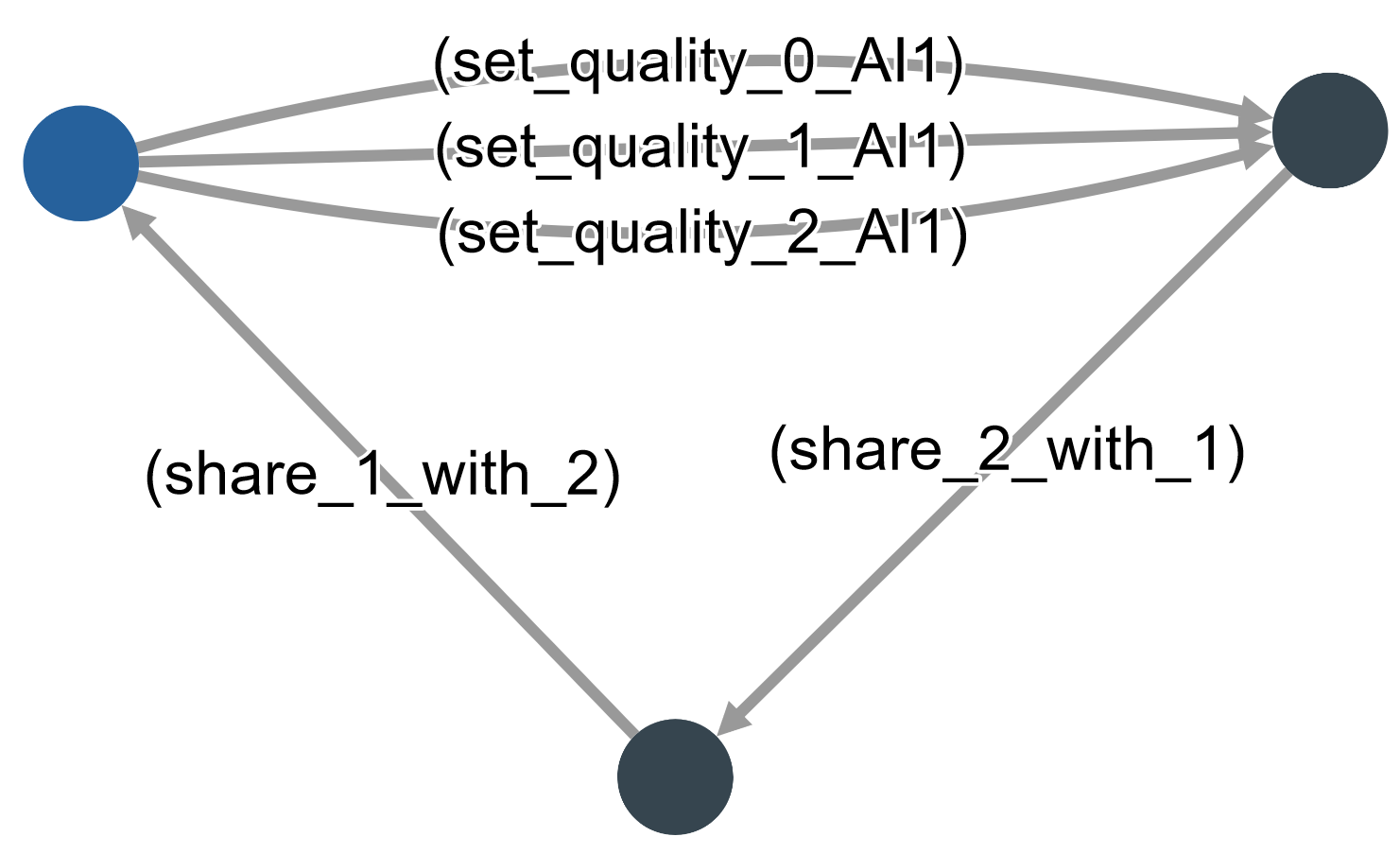}
\caption{Graphical representation of Impersonator in STV}
\label{fig:imp-stv}
\end{figure}

\begin{figure*}[!t]
\centering
\includegraphics[width=0.9\textwidth]{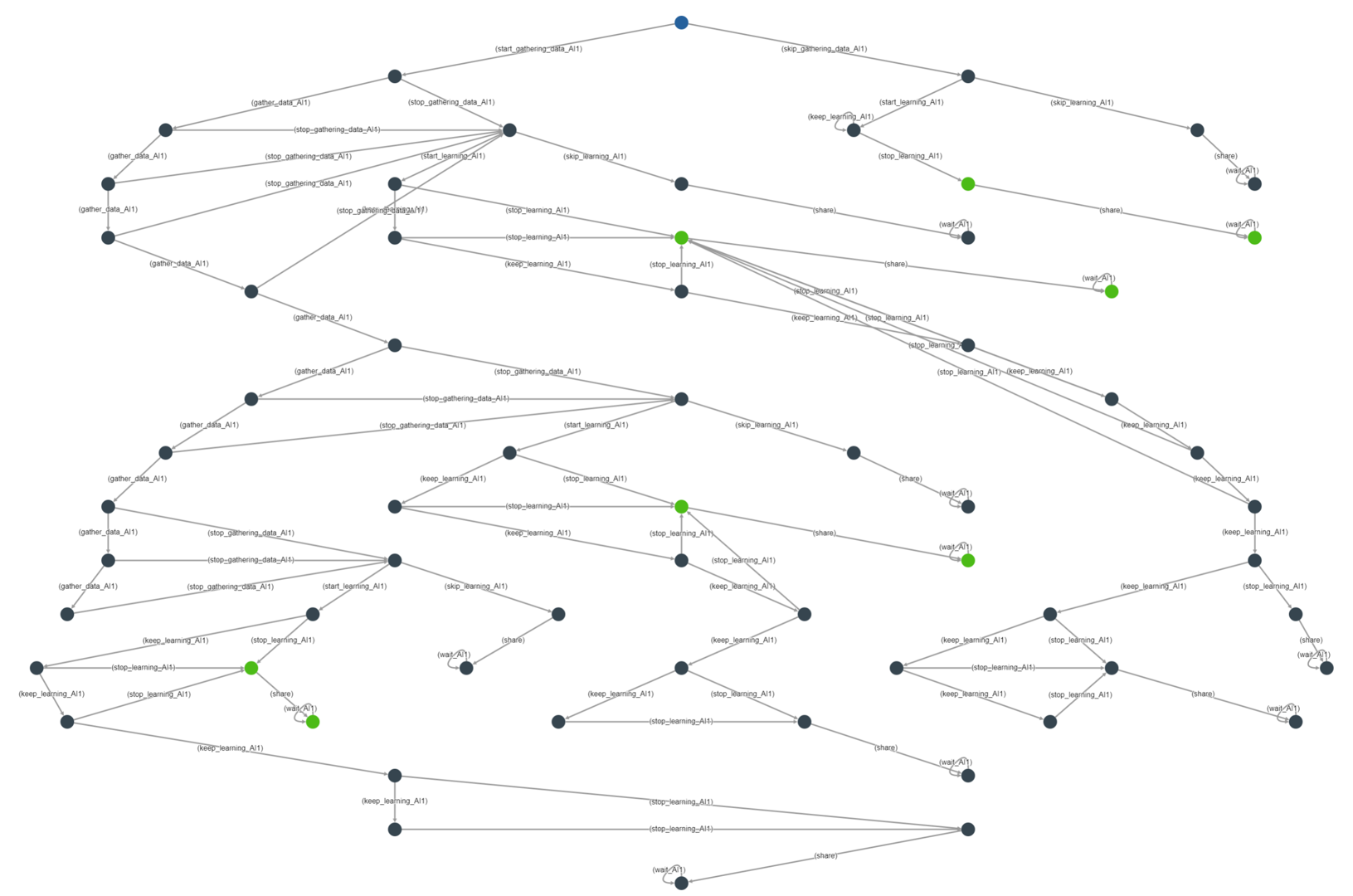}
\caption{Model of \SAI with one honest agent}
\label{fig:model:one}
\end{figure*}

\begin{figure*}[!t]
\hspace{-1cm}\includegraphics[width=1.1\textwidth]{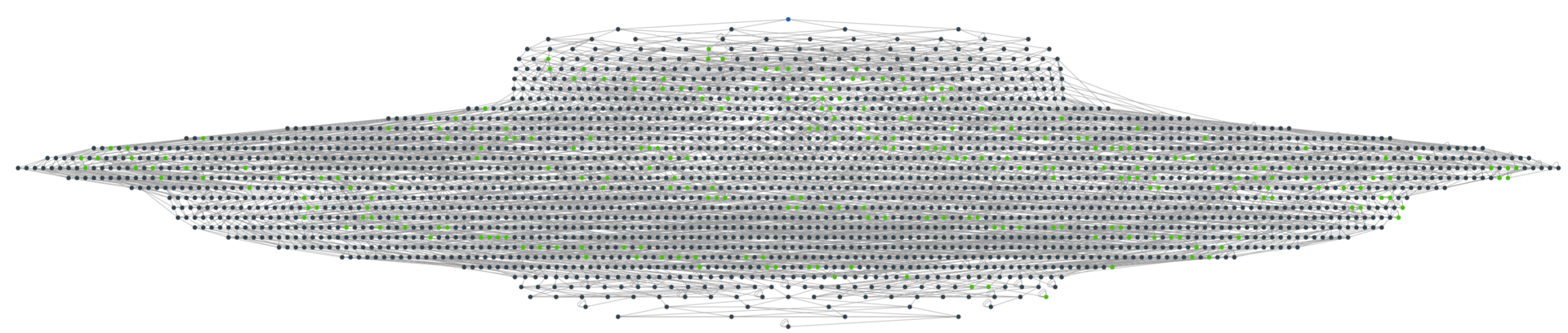}
\caption{Model of \SAI with two honest agents}
\label{fig:model:ufo}
\end{figure*}

\para{Impersonator.}
In this scenario, one of the AI agents is infected with malicious code that results in unwanted behavior.
The agent cannot participate in the data gathering or learning phases, but can share his model with others following the sharing protocol.
The difference between the honest agent and the impersonator is that the latter can fake the quality of his local AI model, hence tricking the next agent into accepting it.
The STV code and its visualization for Impersonator are presented in Figures~\ref{fig:imp} and~\ref{fig:imp-stv}.

\section{\uppercase{Experiments}}\label{sec:experiments}

The STV tool can be used to combine the modules presented in Figures~\ref{fig:honest}--\ref{fig:imp-stv}, and generate the global model of interaction.
We present the output in Figures~\ref{fig:model:one} (for the system with only one honest AI agent) and~\ref{fig:model:ufo} (for two honest agents).
The models can provide invaluable insights into the structure of possible interactions in the system.
Still, a detailed visual scrutiny is possible only in the simplest cases due to the state-space and transition-space explosion.

In more complex instances, we can use STV to attempt an automated verification of strategic-temporal requirements. Since model checking of strategic ability is hard for scenarios with partial observability (\NP-hard to undecidable, depending on the precise syntax and semantics of the specification language), exact verification is infeasible. Instead, we use the technique of \emph{fixpoint approximation}, proposed in~\cite{Jamroga19fixpApprox-aij}, and implemented in STV.
In what follows, we summarise the experimental results obtained that way.

\para{Models and formulas.}
The scalable class of models has been described in detail in Section~\ref{sec:models}.
In the model checking experiments, have used two variants of the system specification, one with a possible impersonation attack, and the other one with the possibility of a man-in-the-middle attack.
In each case, we verified the following formulas:
\begin{itemize}
    \item $\phi_1 \equiv \coop{I}\Always(shared_p\rightarrow(\bigwedge_{i\in[1,n]}mqual_i\leq k))$
    \item $\phi_2 \equiv \coop{I}\Always(shared_p\rightarrow(\bigvee_{i\in[1,n]}mqual_i\leq k))$
\end{itemize}
Formula $\phi_1$ checks whether the Intruder has a strategy to ensure that all honest agents will not achieve quality greater than $k$, while the formula $\phi_2$ checks whether the same is possible for at least one agent.

\para{Configuration of the experiments.}
The experiments have been conducted with the latest version of STV~\cite{Kurpiewski22stv-github}.
The test platform was a server equipped with ninety-six 2.40 GHz Intel Xeon Platinum 8260 CPUs, 991 GB RAM, and 64-bit Linux. 

\para{Results.}
We present the verification results in Figures~\ref{tab:res1} and~\ref{tab:res2}.
\textbf{\#Ag} specifies the scalability factor, namely the number of agents in the system.
\textbf{\#st} and \textbf{\#tr} report the number of global states and transitions in the resulting model of the system, and
\textbf{Gen.} gives the time of model generation.
\textbf{Verif. $\phi_1$} and \textbf{Verif. $\phi_2$} present the verification time and its output for formulas $\phi_1$ and $\phi_2$, respectively.
All times are given in seconds.
The timeout was set to 8 hours.

\para{Discussion of the results.}
In the experiments, we were able to verify models of \SAI for up to 5 agents. 
The verification outcome was conclusive in all cases, i.e., the model checker always returned a concrete answer (True or False).
This means that we successfully model-checked systems for up to almost a billion transitions, which is a serious achievement for an \NP-hard verification problem.
In all cases, formula $\phi_1$ turned out to be false. That is, both impersonation and man-in-the-middle attacks can disrupt the learning process and prevent some agents from obtaining good quality PAIVs.
At the same time, $\phi_2$ was true in all cases. Thus, the design guarantees that at least one good quality PAIV will be achieved, even in the presence of the attacks.

\begin{figure}[t]
    \begin{adjustbox}{width=1.0\columnwidth,center}
    \begin{tabular}{|c|c|c|c|c|c|c|c|c|}
    \hline
    \textbf{\#Ag}               & \textbf{\#st}   & \textbf{\#tr}   & \textbf{Gen.}   & \textbf{Verif. $\phi_1$}   & \textbf{Verif. $\phi_2$}  \\ \hline
    $2$                         & $886$           & $2007$          & $<0.1$          & $<0.1$/False                 & $<0.1$/True            \\ \hline
    $3$                         & $79806$         & $273548$        & $28$            & $151$/False                 & $202$/True           \\ \hline
    $4$                         & $6538103$       & $29471247$      & $1284$          & $5061$/False                & $5102$/True           \\ \hline
    $5$                         & $93581930$      & $623680431$     & $7845$          & $25828$/False               & $25916$/True           \\ \hline
    $6$                         & \multicolumn{5}{c|}{timeout}                                             \\ \hline
    \end{tabular}
    \end{adjustbox}
\vspace{0.1cm}
\caption{Verification results for the Impersonator attack scenario}
\label{tab:res1}
\end{figure}

\begin{figure}[t]
    \begin{adjustbox}{width=1.0\columnwidth,center}
    \begin{tabular}{|c|c|c|c|c|c|c|c|c|}
    \hline
    \textbf{\#Ag}               & \textbf{\#st}   & \textbf{\#tr}  & \textbf{Gen.} & \textbf{Verif. $\phi_1$} & \textbf{Verif. $\phi_2$}  \\ \hline
    $3$                         & $23966$         & $67666$        & $12$          & $21$/False               & $33$/True           \\ \hline
    $4$                         & $4798302$       & $20257664$     & $875$         & $3810$/False             & $3882$/True           \\ \hline
    $5$                         & $71529973$      & $503249452$    & $5688$        & $19074$/False            & $20103$/True           \\ \hline
    $6$                         & \multicolumn{5}{c|}{timeout}                                             \\ \hline
    \end{tabular}
    \end{adjustbox}
\vspace{0.1cm}
\caption{Verification results for Man in the Middle}
\label{tab:res2}
\end{figure}

\section{\uppercase{Conclusions}}\label{sec:conclusion}

In this paper, we present our work in progress on formal analysis of Social Explainable AI. 
We propose that formal methods for multi-agent systems provide a good framework for multifaceted analysis of \SAI environments. 
As a proof of concept, we demonstrate simple multi-agent models of \SAI, prepared with the model checker STV.
Then, we use STV to formalize and verify two variants of resistance to impersonation and man-in-the-middle attacks, with very promising results.
Notably, we have been able to successfully model-check systems for up to almost a billion transitions -- a considerable achievement for an \NP-hard verification problem.

\section*{\uppercase{Acknowledgements}}

The work was supported by NCBR Poland and FNR Luxembourg under the PolLux/FNR-CORE project STV (POLLUX-VII/1/2019), and the CHIST-ERA grant CHIST-ERA-19-XAI-010 by NCN Poland (2020/02/Y/ST6/00064).
The work of Damian Kurpiewski and Teofil Sidoruk was also supported by the CNRS IEA project MoSART.

\bibliographystyle{plain}
{\small
\bibliography{wojtek,wojtek-own}}

\end{document}